\documentstyle[twocolumn,aps,psfig]{revtex}

\begin{document}
\draft
\flushbottom
\twocolumn[
\hsize\textwidth\columnwidth\hsize\csname @twocolumnfalse\endcsname

\title{ Single-photon tunneling}
\author{I.I. Smolyaninov$^{(1)}$, A.V. Zayats$^{(2)}$, A. Gungor$^{(3)}$, C.C. Davis $^{(1)}$ }
\address{ $^{(1)}$ Electrical and Computer Engineering Department,
University of Maryland, College Park, MD 20742}
\address{$^{(2)}$ Department of Pure and Applied Physics,
Queen's University of Belfast, Belfast BT7 1NN,UK}
\address{ $^{(3)}$Department of Physics,
Fatih University, Istanbul, Turkey}

\date{\today}
\maketitle
\tightenlines
\widetext
\advance\leftskip by 57pt
\advance\rightskip by 57pt

\begin{abstract}
Strong evidence of a single-photon tunneling effect, a direct analog of single-electron tunneling, has been obtained in the measurements of light tunneling through individual subwavelength pinholes in a thick gold film covered with a layer of polydiacetylene. The transmission of some pinholes reached saturation because of the optical nonlinearity of polydiacetylene at a very low light intensity of a few thousands photons per second. This result is explained theoretically in terms of "photon blockade", similar to the Coulomb blockade phenomenon observed in single-electron tunneling experiments. The single-photon tunneling effect may find many applications in the emerging fields of quantum communication and information processing. 


\end{abstract}

]
\narrowtext

\tightenlines

The emerging fields of quantum communication and information processing require unusual sources of light with strong quantum correlations between single photons, and new types of light detectors that can detect individual photons without destroying them \cite{1}. A few such novel quantum devices have been demonstrated very recently, such as a quantum dot single-photon turnstile device, and a quantum non-demolition photon detector based on strong photon-matter coupling \cite{2,3}. Operation of such quantum devices is extremely involved, and there is a great need for novel ideas and device concepts in this field. One such idea is the concept of photon blockade in a nonlinear optical cavity \cite{4}, introduced theoretically in close analogy with the well-known phenomenon of Coulomb blockade for quantum-well electrons. The suggested realization of this concept involves a four-level atomic system in a macroscopic optical cavity, which exhibits resonantly enhanced two-photon absorption limited Kerr nonlinearity. The optical transmission of such a cavity is supposed to exhibit photon anti-bunching. 

In this paper we suggest and demonstrate experimentally another example of photon blockade, which is based on close analogy between electron and photon tunneling. It is well known that Coulomb blockade leads to the observation of single-electron tunneling in tunnel junctions with an extremely small capacitance, where the charging energy $e^2/2C$ of the capacitance $C$ is much larger than the thermal energy $k_BT$ and the quantum fluctuation energy $\hbar /\tau $ with $\tau =RC$ (where $R$ is the resistance of the tunnel junction) \cite{5,6}. An example of such a tunnel junction is shown in Fig.1(a), where a nanoparticle with a very small capacitance $C$ is placed in the gap between the tip and sample of a scanning tunneling microscope (STM). In this case, tunneling of a single electron into the nanoparticle results in noticeable charging of the junction capacitance, so the probability of other tunneling events is drastically reduced. Therefore there is a strong correlation between tunneling events \cite{5,6}: electrons tunnel one at a time, and steps ("Coulomb staircase") in the current-voltage characteristic of a tunnel junction are observed. 
\begin{figure}[tbp]
\centerline{
\psfig{figure=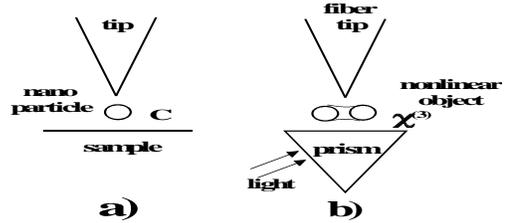,width=8.5cm,height=5.0cm,clip=}
}
\caption{ Schematic views of single-electron (a) and single-photon (b) tunneling experiments. The nonlinear object in (b) is represented by two metallic nanoparticles separated by a gap filled with nonlinear optical material. 
}
\label{fig1}
\end{figure}

Let us consider a geometry of an optical single-photon tunneling experiment Fig.1(b) designed to emulate the geometry of a single-electron tunneling experiment in Fig.1(a). Classical realization of light tunneling is based on a glass surface illuminated in the total internal reflection geometry, e.g., using a prism or a semicylinder. In this case, all incident light is reflected and only an evanescent field (exponentially decaying from the surface) exists over a smooth surface. If a tapered glass fiber is placed sufficiently close to the glass-air interface, the evanescent field is transformed into propagating waves in the fiber. Thus, optical tunneling through an air gap (which can be considered as a tunnel barrier) occurs. This geometry is used in scanning tunneling optical microscope (STOM) operation.  

To complete our analogy to single-electron tunneling, we will consider a nanometer-scale object in the tunnel gap of a STOM. Let us assume that this object posses nonlinear optical properties and exhibits well defined localized electromagnetic modes. There are quite a few examples of such objects. It may be a silver-coated polymerized diacetylene composite nanoparticle \cite{7}, or a narrow gap between two silver particles filled with nonlinear optical material. Nanometer-size gaps between metal surfaces (such as a tip and a sample of an STM) are known to exhibit pronounced and well-defined localized surface plasmon (LSP) resonances \cite{8,9}. Excitation of LSP modes may lead to very large local electromagnetic field intensity enhancement \cite{10} because of the very small volume of these modes. If the frequency of tunneling photons is in resonance with some localized optical mode of our nanometer-scale object in Fig.1(b), tunneling from the sample into the tip of the STOM is facilitated. This happens via excitation of the localized mode. The electric field of the excited localized mode $E_L$ induces local changes in the dielectric constant $\epsilon $ of the nonlinear optical material:
\begin{equation} 
\epsilon = \epsilon ^{(1)} + 4\pi \chi ^{(3)}E^2_L ,
\end{equation}
where $\epsilon ^{(1)}$ and $\chi ^{(3)}$ are the linear dielectric constant and the third order nonlinear susceptibility, respectively. As a consequence of dielectric constant change, the localized modes may experience a noticeable frequency change, so that the tunneling photons will not be in resonance any more. Thus, photon tunneling will be blocked in a manner similar to the Coulomb blockade effect for electrons in Fig.1(a), until the localized optical mode has radiated and the nonlinear material relaxes into its initial state. Our calculations below show that in some cases even tunneling of a single optical quantum into the localized plasmon mode is enough to cause an effective blockade of the tunneling gap for the next optical quanta. 

Unless we are interested in the space-time dynamics of photon tunneling, a formal analogy between electron and photon tunneling can be drawn \cite{11}. In the widely used theory of STM \cite{12} the tunneling current is proportional to the local density of states (LDOS) of a sample under investigation,  $\rho _s({\mathbf r}, E_F )$, at the Fermi energy:
\begin{equation}
I \sim \rho_s({\mathbf r},E_F),
\end{equation}
This theory adequately describes the resolution of the microscope, gives the interpretation of the images and explains some influences of the tip. Single-electron tunneling can be realized in systems in which tunneling electrons (ultimately, one electron) significantly modify the energy spectrum of LDOS $\rho _s({\mathbf r}, E_F )$ by, for example, influencing the Fermi energy of the sample (the nanoparticle in Fig. 1(a)).
\begin{figure}[tbp]
\centerline{
\psfig{figure=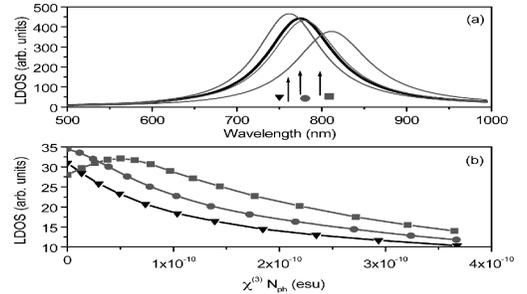,width=9.0cm,height=6.0cm,clip=}
}
\caption{ 
LDOS spectra (a) and intensity dependencies (b) of LDOS at the wavelengths indicated by arrows calculated for the 10 nm spherical silver particle placed at 1 nm over a silver film coated with a nonlinear material. The LDOS spectra correspond to different changes of the refractive index of a nonlinear material: (from left to right) $\Delta n = -0.03, 0, 0.01$, and 0.1. The linear refractive index of the nonlinear material ($n_0 = 1.7$) is chosen close to the refractive index of BCMU polydiacetylene.
}
\label{fig2}
\end{figure}

The LDOS for photons (sometimes referred as the photonic mode density)
 can be defined in the same way as is done for electrons via the electric field dyadic
 Green function ${\mathbf G}(\mathbf r,\mathbf r',\omega )$ of a system under
 consideration \cite{13}:
\begin{equation}
\rho({\mathbf r}, \hbar \omega )=-1/\pi Im{\mathbf G}({\mathbf r},{\mathbf r},\hbar
 \omega )
\end{equation}
which gives the density of states of electromagnetic eigenmodes of energy $\hbar \omega $. The physical interpretation of the LDOS in optics is that it is directly related to the square of the modulus of the electric field at a given point in space ${\mathbf r}$ and at a given photon energy $\hbar \omega $, and, therefore, to the electromagnetic field enhancement in the system. 

Similar to single-electron tunneling, which is observed in systems in which tunneling electrons significantly modify the energy spectrum of LDOS, tunneling photons can significantly modify the LDOS spectrum of a system exhibiting third-order nonlinear effects through local changes of the dielectric constant. As a consequence of the refractive index change, the polarizability and, therefore, the LDOS of the system are modified. A nonabsorbing electro-optical nonlinearity should be employed to avoid optical losses in a nonlinear material. For a particle made of dielectric material with a low dispersion refractive index, the LDOS has a broad continuous spectrum, thus requiring significant intensities of the incoming light for controlling the particle polarizability. In contrast, small metallic particles exhibit a narrow-band LDOS spectrum in the spectral range where localized surface plasmons are excited. In the maximum of the band, a significant enhancement of the electromagnetic field occurs in the vicinity of the particle in turn enhancing the nonlinear effects. Thus, realization of light controlled photon tunneling can be achieved combining high third-order nonlinear materials with good metals exhibiting a narrow spectrum of LDOS and strong field enhancement effects. 

A very strong field enhancement is expected when a metal sphere is placed close to a metal surface so that the electromagnetic near-field interaction modifies significantly the plasma resonance of the sphere resulting in the surface plasmon's localization at the sphere-surface junction. As the highest field enhancement is expected in this system, in general, for small sphere-surface distances $d$, we shall consider the situation $d<<R$ that simplifies the LDOS spectrum calculation significantly \cite{8,9}. This model has been successfully employed for describing the surface enhanced Raman scattering, photon emission from STM, etc. A similar model, considering spherical particles at small separation distance, describes giant enhancement of the electromagnetic field and related optical processes in self-affine structures and fractals \cite{10}. In our case we shall consider a metal sphere placed over a metal surface and a nonlinear material is assumed to fill the sphere-surface gap. The LDOS spectrum of such a system can be calculated analytically in the limit $d<<R$ since it formally resembles particle motion in the Coulomb field \cite{14}, and is determined by the following dispersion relation \cite{8}
\begin{equation}
Re(\epsilon /\epsilon(\omega ))  = (m+1/2)(d/2R)^{1/2},  m=0, 1, 2
... ,
\end{equation}
where $\epsilon $ and $\epsilon (\omega )$ are the dielectric constants of nonlinear material and metal, respectively, $R$  is the radius of the sphere, and $d$ is the distance between sphere and surface. The quantum number $m$ corresponds to different localized surface plasmon modes at the sphere-surface junction. The strongest localization occurs for the lowest, $m = 0$, mode with a localization length  $L=(2dR)^{1/2}$ , and the electric field of this localized mode is determined by the potential
\begin{equation}
\phi(k)  = A \frac{exp(-kL)}{k\epsilon },
\end{equation}
where A is the normalization constant. Having determined the mode volume of this localized plasmon, the respective electric field in the gap can be calculated and the related electro-optical nonlinearity estimated (Eq. 1) for a given number of tunneling photons \cite{15}. The results of our calculations are shown in Fig. 2. The LDOS in the gap is presented as a function of light wavelength (Fig.2a) for different values of the nonlinear refractive index change caused by excitation of a localized plasmon mode. Fig.2(b) shows the changes in the LDOS at different light intensities (different numbers of plasmon quanta excited in the gap) for the incident light wavelength corresponding to the LSP resonance at low light intensities, as well as for the longer and shorter wavelengths. With an increase of the incident light intensity the LDOS resonance shift leads to significant variations of LDOS at the illumination wavelength. The decrease of the LDOS results in the saturation of transmitted light intensity with an increase of the incident light intensity. Because of very strong field localization, the number of photons required to achieve a "photon blockade" is very low. If  $\chi ^{(3)}\sim 10^{-10}$ esu is assumed, which is on the correct order of magnitude for the $\chi ^{(3)}$ observed in poly-3-butoxy-carbonyl-methyl-urethane (3BCMU) and 4BCMU polydiacetylene materials \cite{16} (these materials hold the record for the largest fast nonresonant optical nonlinearity), there is a fair chance to observe single-photon tunneling, since a single tunneled photon causes the reduction of LDOS at the incident light wavelength by almost a factor of 2. In this scenario, the photon which is close to the resonance of the LDOS spectrum tunnels through a barrier via a localized sphere-surface plasmon. The electric field of this LSP acting on the nonlinear material results in a shift of the LDOS spectrum according to the refractive index change, thus blocking tunneling of subsequent photons. Only when the LSP is radiated and after the nonlinear material is relaxed into its initial state, the LDOS resonance at this wavelength is recovered, and another photon can tunnel. Therefore, the temporal behavior of the tunneling is governed by the time of nonlinear material thermalization and the LSP lifetime.

\begin{figure}[tbp]
\centerline{
\psfig{figure=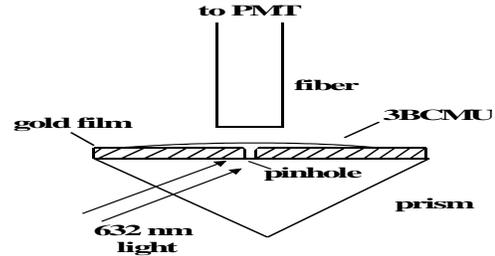,width=9.0cm,height=5.0cm,clip=}
}
\caption{ Schematic view of our experimental setup. 
}
\label{fig3}
\end{figure}

Our measurements of light tunneling through individual subwavelength pinholes in a thick gold film covered with a layer of polydiacetylene provide strong evidence of single-photon tunneling. Our experimental setup is shown in Fig.3. A thick ($\sim 1 \mu m$ ) gold film had been deposited on the surface of a glass prism. Only a few sparsely separated pinholes were visible in the film under illumination with 632 nm light at an angle larger than the angle of total internal reflection for the glass-air interface. A drop of 3BCMU polydiacetylene solution in chloroform \cite{17} was deposited onto the gold film surface. After solvent evaporation a thick film of polydiacetylene was left on the surface. A cleaved optical fiber with a core diameter of about $9 \mu m$ was used to collect 632 nm light transmitted through the individual pinholes. The fiber was positioned above individual pinholes at a distance of a few micrometers using a far-field optical microscope and the shear-force distance control system commonly used in near-field optical setups \cite{18}. The pinholes selected for the measurements had very low optical transmission of a few hundred photons per second. Fig.4 shows the measured dependencies of the transmitted light as a function of the illuminated light intensity for some selected pinholes. The size $a$ of these pinholes may be roughly estimated using the Bethe's expression for the cross section of a subwavelength aperture \cite{19}: $A\sim a^2(a/\lambda )^4$. Using the numerical data from Fig.4(a) we have got an estimate of $a \sim 7$ nm, which corresponds to a desired range of pinhole sizes. While the dependencies obtained for most pinholes were linear, transmission of some small pinholes exhibited saturation and even staircase-like behavior. 
\begin{figure}[tbp]
\centerline{
\psfig{figure=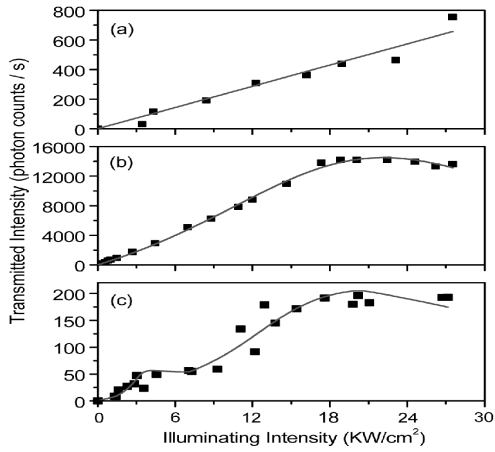,width=9.0cm,height=7.0cm,clip=}
}
\caption{ Experimentally measured transmission of selected sub-wavelength pinholes at 632 nm. Saturation of transmission due to the optical nonlinearity of polydiacetylene at a very low light intensity of a few thousands photons per second has been observed. Solid lines are the fits in the model of the intensity dependent LDOS resonances.
}
\label{fig4}
\end{figure}

The experimental data has been fitted assuming the intensity dependent LDOS resonances analogous to the model described above. Although the absolute spectral positions of the resonances related to nanopores are somewhat different from the ones given by equation (4), the overall behavior of localized plasmon modes in nanopores is expected to be similar \cite{8}. The fitting curves obtained with one (Fig. 4b) and two (Fig. 4c) intensity dependent local plasmon resonances show good agreement with the experiment. 
The $\delta \lambda /\lambda $ of the resonances obtained from the fit was equal to 0.05 and 0.3 for low and high intensity resonances in Fig. 4c, respectively, and 0.6 for the resonance in Fig. 4b. These values have a similar order of magnitude to those observed in small metal particles \cite{10}. For a two-resonance curve (Fig.4c), the second resonance observed at higher light intensity is presumed to be initially far away from the frequency of the incident light (c.f. Fig. 2a). It becomes significant at higher intensities leading to the staircase-like behavior of the observed  transmission.

In conclusion, strong evidence of a single-photon tunneling effect, a direct analog of single-electron tunneling, has been obtained in measurements of light tunneling through individual subwavelength pinholes in a thick gold film covered with a layer of polydiacetylene. Single-photon tunneling effect may find many applications in the emerging fields of quantum communication and information processing.

\end{document}